\documentclass[ALICE,manyauthors]{ALICE_Package/cernphprep}
\usepackage[comma,square,numbers,sort&compress]{natbib}
\usepackage{hyperref}
\usepackage{lineno}
\usepackage[utf8]{inputenc}
\usepackage{amsmath}
\usepackage{amsfonts}
\usepackage{amssymb}
\usepackage{amsthm}
\usepackage{bm}
\usepackage[english]{babel}
\usepackage[T1]{fontenc}
\usepackage{graphicx}
\usepackage{enumerate}
\usepackage{textcomp}
\usepackage[toc,page]{appendix}
\usepackage{slashed}
\usepackage{lineno}
\usepackage{color}

\usepackage{setspace}
\graphicspath{ {figure/}}

\newcommand{ \la }{\langle}
\newcommand{ \ra }{\rangle}
\def \mean#1 {{\la #1 \ra}}

\begin{document}

\begin{titlepage}
\PHyear{2021}
\PHnumber{199}      
\PHdate{30 September}  
\title{\texorpdfstring{
General balance functions of identified charged hadron pairs of $\mathbf{(\mathbf{\pi},K,p)}$ in Pb--Pb collisions at $\mathbf{\sqrt{\textit{s}_{_{\rm \mathbf{NN}}}}} =$ 2.76 TeV}{}}
\ShortTitle{General Balance Functions} 
\Collaboration{ALICE Collaboration\thanks{See Appendix~\ref{app:collab} for the list of collaboration members}}
\ShortAuthor{ALICE Collaboration} 

\begin{abstract}
First measurements of  balance functions (BFs) of all combinations of identified charged hadron $(\pi,\rm K,\rm p)$ pairs in Pb--Pb collisions at $\sqrt{s_{_{\rm NN}}} = 2.76$ TeV recorded by the ALICE detector are presented. The BF measurements are carried out as two-dimensional differential correlators versus the relative rapidity ($\Delta y$) and azimuthal angle ($\Delta\varphi$) of hadron pairs, and studied as a function of collision centrality. The $\Delta\varphi$ dependence of BFs is expected to be sensitive to the light quark diffusivity in the quark--gluon plasma. While the BF azimuthal widths of all pairs substantially decrease from peripheral to central collisions, the longitudinal widths exhibit mixed behaviors: BFs of $\pi\pi$ and cross-species pairs narrow significantly in more central collisions, whereas those of $\rm KK$ and $\rm pp$ are found to be independent of  collision centrality. This dichotomy is qualitatively consistent with the presence of strong radial flow effects and the existence of two stages of  quark production in relativistic heavy-ion collisions.  
Finally, the first measurements of the collision centrality evolution of BF integrals are presented, with the observation that   charge balancing fractions are nearly independent of  collision centrality in Pb--Pb collisions. Overall, the results presented provide new and challenging constraints for theoretical models of hadron production and transport in relativistic heavy-ion collisions.
\end{abstract}
\end{titlepage}

\setcounter{page}{2}

Convincing evidence for the production of strongly interacting quark--gluon plasma (QGP) in  heavy-ion (AA) collisions has been reported from a variety of measurements at the Relativistic Heavy Ion Collider (RHIC) and the Large Hadron Collider (LHC)~\cite{Adams:2005dq,Adcox:2004mh,Arsene20051,Back:2004je}, including observations of  strong elliptic flow~\cite{Adams:2004bi,Heinz:2013th,PhysRevLett.105.252302}, suppression of high transverse momentum ($p_{\rm T}$) hadron production~
\cite{Adams:2003im,
ALICE:2019hno,
PhysRevLett.91.172302,
PhysRevC.69.034910,
PhysRevLett.97.152301,
ALICE:2015mjv}, suppression of quarkonium states\cite{ALICE:2013osk,CMS:2017uuv,CMS:2018zza,ALICE:2020vjy,ALICE:2019nrq,ALICE:2016flj},  as well as dihadron correlation  functions~\cite{Abelev:2009ah,ALICE:2011gpa}.
Many of these findings are quantitatively explained by hydrodynamic calculations in which the QGP matter undergoes radial and azimuthally anisotropic collective motion.
The existence of the latter is well established based on measurements of flow coefficients with finite pseudorapidity ($\eta$) gap and multi-particle cumulants, whereas the presence of the former is inferred in part from the   increase of average transverse momenta  with the mass of  hadrons~\cite{ALICE:2013Pt}, the centrality dependence of  event-by-event $p_{\rm T}$ fluctuations~\cite{Abelev:2014ckr,Basu:2016ibk}, as well as the observed narrowing of the near-side peak of balance functions (BFs) in central collisions relative to that observed in peripheral collisions~\cite{Adams:2003kg,Aggarwal:2010ya,Adamczyk:2015yga,Abelev:2013csa,Adam:2015gda,Basu:2020ldt}. Balance functions essentially amount  to differences of  correlation functions of like-sign and unlike-sign charges.
They are measured, typically, as functions of particle pair separation in azimuth angle and rapidity. They indicate the degree to which the production of a positive charge is accompanied by the production of a negative charge somewhere in phase space. As such, BFs probe the balancing of charge distributions in momentum space and  theoretical studies show they are sensitive to the details of the time (i.e., whether particles are produced early or late), production mechanisms, and transport of balancing charges. 

Measurements of BFs were originally proposed as a tool to investigate the  delayed hadronization   and two stages of  quark production in the  QGP formed in AA collisions~\cite{Bass:2000BF}. These terms refer to the notion that quark production occurs in two distinct stages, the first at the onset, and the second at the very end (just before hadronization and freeze-out) of AA collisions. The two stages are posited to be separated by a period of isentropic expansion whose duration depends on the multiplicity of produced quarks and gluons and thus the collision impact parameter. Hadron pairs produced at the onset of collisions feature large longitudinal separation (i.e., rapidity differences $\Delta y$) whereas pairs produced after the expansion   have smaller $\Delta y$ separations determined by the smaller temperature of the system at that time. AA collisions with smaller impact parameters are expected to produce larger systems with a longer isentropic stage in which late particle production dominates. The longitudinal and azimuthal widths of BFs are thus expected to progressively decrease from peripheral to central collisions as the fraction of late particle production 
increases. BFs could also provide a  precise probe of balancing particle production~\cite{Jeon:2002BFCF, Pratt:2002BFLH, Pratt:2015PC, Bialas:2004BFCO}, the hadrochemistry of particle production~\cite{Pratt:2015PC,Pratt:2018ebf},  as well as the collision dynamics~\cite{Voloshin:2006TRE,Bozek:2005BFTF}. Recent studies also indicate that the  BF dependence on pair separation in azimuth is sensitive to  the diffusivity of light quarks, a measure of the diffusion and scattering of quarks within the QGP, which  has thus  far received only limited attention~\cite{Pratt:2019pnd, Pratt:2018ebf}.
Finally, BFs also provide a tool to calibrate measurements of  the Chiral Magnetic Effect~\cite{Ye:2019EMBF,Alam:2018owe} and net charge/baryon  fluctuations deemed essential for the determination of QGP susceptibilities~\cite{Karsch:2010ck,Pruneau:2019BNC}.

Few  measurements of BFs of identified hadrons have been reported to date.  At RHIC, these include BF measurements of charged hadrons, pion pairs, kaon pairs, as well as proton/antiproton pairs~\cite{Adams:2003kg,Aggarwal:2010ya,Adamczyk:2015yga}, whereas at the LHC, only  charged hadron BFs have been reported~\cite{Abelev:2013csa,Adam:2015gda}.  Of these, only the results published by ALICE were fully corrected for detector acceptance and particle losses (efficiency). Integrals of measured BFs have not been considered and no cross-species BFs have been published. Theoretical analyses of measured BFs have consequently  focused mainly on the interpretation of the narrowing with collision centrality of charged hadron BFs.
The full potential of BFs as a probe of the evolution dynamics and chemistry of the QGP has thus so far been underexploited.  In this paper, general balance functions  of identified charged hadron species  $(\pi,\rm K,\rm p)$ are reported for the first time. These general BFs are corrected for efficiency and non uniform acceptance effects and it becomes possible to study the effects of two-stage quark production, light quark diffusivity, and relative balancing fractions 
using BFs of nine distinct identified pairs of charged hadron species.

 The BF of a species of interest, $\alpha$, and an associated species, $\beta$, was originally defined in terms of conditional densities~\cite{Bass:2000BF} but it is convenient  to compute BFs in terms of normalized cumulants $R_2$ according to
\begin{equation}
\begin{aligned}
B^{\alpha\beta}(\vec{p}_{\alpha},\vec{p}_{\beta}) = \frac{1}{2} 
 & \left\{
 \rho_1^{\beta^-}(\vec{p}_{\beta^-})
 \left[
 R_2^{\alpha^+\beta^-}(\vec{p}_{\alpha^+},\vec{p}_{\beta^-}) - {R_2^{\alpha^-\beta^-}(\vec{p}_{\alpha^-},\vec{p}_{\beta^-})} 
 \right] +\right. \\
 & \left.\;\;\;
\rho_1^{\beta^+}(\vec{p}_{\beta^+})
\left[ 
R_2^{\alpha^-\beta^+}(\vec{p}_{\alpha^-},\vec{p}_{\beta^+}) - {R_2^{\alpha^+\beta^+}(\vec{p}_{\alpha^+},\vec{p}_{\beta^+})} 
\right] 
\right\}
\end{aligned},
\label{eq:BFR2}
\end{equation}
with
\begin{equation}
\label{eq:R2}
R_2^{\alpha\beta}(\vec{p}_{\alpha},\vec{p}_{\beta}) \equiv \frac{\rho_2^{\alpha\beta}(\vec{p}_{\alpha},\vec{p}_{\beta})}{{\rho_1^{\alpha}(\vec{p}_{\alpha})}{\rho_1^{\beta}(\vec{p}_{\beta})}} - 1
\end{equation}
where $\rho_1^{\alpha}(\vec{p}_{\alpha})\equiv {\rm d}N/{\rm d}\vec p_{\alpha}$ and $\rho_2^{\alpha\beta}(\vec{p}_{\alpha},\vec{p}_{\beta})\equiv {\rm d}N_{\rm pair}/{\rm d}\vec p_{\alpha}{\rm d}\vec p_{\beta}$ are single- and particle-pair densities of species $\alpha$ and $\beta$ measured at momenta $\vec{p}_{\alpha}$ and $\vec{p}_{\beta}$, respectively, while labels $+$ and $-$ stand for positive and negative charges. Normalized cumulants  $R_2$ are robust observables, i.e., independent  to first order of measurement efficiencies. They are   sensitive to the strength of correlation between species $\alpha$ and $\beta$. Their properties were described in several  publications~\cite{PhysRevC.100.044903,Adam:2017ucq,Ravan:2013lwa,Gonzalez:2018cty}. The  combination of $R_2$ correlation functions, normalized by single particle densities, as per Eq.~(\ref{eq:BFR2}), is strictly equivalent to the balance function  introduced in Ref.~\cite{Jeon:2002BFCF,Bass:2000BF} and measures the correlation between positive and negative particles of species $\alpha$ and $\beta$  constrained by charge conservation. Integrals  of inclusive charge balance functions,  $I_{B}^{+-}(\Omega)\equiv \int_{\Omega} B^{+-} {\rm d}\Delta\eta$,  are expected to  lie within the range $0< I_{B}^{+-}(\Omega)\le 1$ for limited acceptances $\Omega$. However, they converge to unity for full acceptance coverage. 
Furthermore,  fractions   $I_{B}^{\alpha\beta}(\Omega)/I_{B}^{\alpha}(\Omega)$ are determined by the hadrochemistry of the QGP and transport properties of the medium.  In the full acceptance coverage limit, the denominator of this fraction must satisfy $I_{B}^{\alpha}(\Omega) \equiv \sum_{\beta} I_{B}^{\alpha\beta}(\Omega)\rightarrow 1$~\cite{Pruneau:2019BNC}.

In this paper, the identified particle BFs of  nine pairs of charged hadrons ($\pi^{\pm}$, ${\rm K}^{\pm}$ and $\rm p/\overline p$) $\otimes$ ($\pi^{\pm}$, ${\rm K}^{\pm}$, and $\rm p/\overline p$) are reported as joint functions of the relative rapidity ($\Delta y$) and azimuthal angle ($\Delta \varphi$) and studied as a function of collision centrality. 
Measurements of $R_2^{\alpha\beta}(\vec{p}_{\alpha},\vec{p}_{\beta})$ are carried out  in terms of the rapidity  and azimuthal angle $y_{\alpha}$,  $\varphi_{\alpha}$,  $y_{\beta}$, and $\varphi_{\beta}$ for   fixed $p_{\rm T}$ ranges, and averaged across the pair acceptance  to yield correlation functions $R_2^{\alpha\beta}(\Delta y,\Delta\varphi)$ with   $\Delta y=y_{\alpha}-y_{\beta}$ and $\Delta\varphi =\varphi_{\alpha}-\varphi_{\beta}$ following  the procedure used in Ref.~\cite{PhysRevC.100.044903}. The  densities of associated particles, $\rho_1^{\beta}$, used in Eq.~(\ref{eq:BFR2}), are integrated  from $p_{\rm T}$-dependent densities reported in prior ALICE measurements~\cite{ALICE:2013Pt} to match the  $p_{\rm T}$ ranges used in measurements of the normalized cumulants $R_2$. The correlators $R_2^{\alpha\beta}$ and densities $\rho_1^{\beta}$  are corrected for $p_{\rm T}$-dependent particle losses and non uniform acceptance. Densities $\rho_1^{\beta}$ were additionally corrected for minor contamination effects as per the procedure described in ~\cite{ALICE:2013Pt}. The measured BFs thus feature  absolute normalization which enables meaningful determination of their integrals and collision centrality dependence.  

As already mentioned, the shape of the BFs vs. $\Delta y$ and $\Delta\varphi$ is sensitive to the timescales at which particles are produced during the system evolution. Early emission occurs at large effective collisional energy  $\sqrt{s}$ and  is thus expected to yield broad BFs in $\Delta y$ and $\Delta\varphi$, whereas late emission, at collisional energy commensurate with the system temperature, is expected to produce much narrower
near side peak correlations vs. $\Delta y$ and $\Delta\varphi$~\cite{Bass:2000BF}. Additionally, the integral of the BFs shall also provide increased sensitivity to the hadrochemistry of the collisions. Indeed whereas contributions to single-particle spectra from hadronic resonance decays  must be inferred from models, integrals of the BFs are directly sensitive to the magnitude of  (hadronic) feeddown contributions. For instance, by comparing the integrals of $\pi^+\pi^-$ and $\pi^{\pm}{\rm K}^{\mp}$ BFs, sensitivity to the relative strengths of processes that lead to such correlated pairs of particles is acquired. It becomes possible to better probe the role of hadronic resonance decay contributions  and increased sensitivity to the hadrochemistry of the QGP and its susceptibilities is gained~\cite{Pratt:2018ebf}.

The BFs presented are based on $1\times10^{7}$ minimum bias (MB) Pb--Pb collisions at $\sqrt{s_{\rm NN}}$ = 2.76 TeV collected in 2010 by the ALICE collaboration.
Descriptions of the ALICE detector and its performance have been reported elsewhere~\cite{1748-0221-3-08-S08002,Abelev:2014ffa}.
The minimum bias trigger required a combination of hits in the V0 detectors and layers of the SPD detector.
The  V0 detectors, which cover the  full azimuth and the pseudorapidity ranges $-3.7 < \eta < -1.7$ and $2.8 < \eta < 5.1$, also provided a measurement of the charged particle multiplicity used to classify  collisions into centrality classes corresponding to 0--5\% (most central) to 80--90\% (most peripheral) of the Pb--Pb hadronic cross section~\cite{Aamodt:2010cz}. Some centrality classes have been combined to optimize the statistical accuracy of the BFs reported. Particle momenta were determined based on Kalman fits of  charged particle tracks reconstructed in the Time Projection Chamber (TPC).
The particle identification (PID) of charged hadrons was performed based on specific energy loss (${\rm d}E/{\rm d}x$) measured in the TPC and particle velocities  measured in the Time-of-Flight  detector (TOF). Track quality criteria based on the number of space points, the distance of closest approach to the collision primary vertex, and the $\chi^2$ of the Kalman fits were used to restrict the measurements to primary particles produced by the  Pb--Pb collisions and suppress contamination from tracks resulting from weak decays and interactions of particles with the apparatus. Additionally, PID selection criteria based on deviations of ${\rm d}E/{\rm d}x$ and TOF from their respective expectation values, at a given momentum, and for each species of interest, were used to optimize the species purity. These and other selection criteria are reported in  detail below in the context of a discussion  of systematic uncertainties.
The analysis focused on the low $p_{\rm T}$ range, commonly known as the ``bulk" physics regime. Slightly different $p_{\rm T}$ ranges were used for each species to optimize yields and species purity. Charged pions and kaons were selected in the range $0.2\le p_{\rm T} \le 2.0$ GeV/$c$, whereas (anti-)protons are within $0.5\le p_{\rm T} \le 2.5$ GeV/$c$. The selected rapidity range, largely determined by the TOF  coverage, was set to $|y_{\pi}|\le0.8$ and $|y_{\rm p}|\le0.6$ for measurements of  $B^{\pi\pi}$ and $B^{\rm pp}$, respectively, 
 and set to  $|y|\le0.7$ for all other BFs reported.

Track reconstruction efficiencies  and PID purity were studied with Monte Carlo simulations of Pb--Pb collisions produced with the HIJING generator~\cite{Wang:1991hta} and propagated through a model of the ALICE detector with GEANT3~\cite{Brun:1082634}. Selected track quality and PID criteria yield  purities of 97\%, 95\%, and 94\% for $\pi^{\pm}$, ${\rm K}^{\pm}$, and ${\rm p/\overline p}$, respectively, thereby minimizing species contamination and its impact on correlation functions.
Corrections for track losses were carried out using a weighting technique~\cite{Ravan:2013lwa}. Weights are calculated independently for positive and negative tracks of each species considered, for each  centrality range, both magnetic field polarities used in the measurements, versus $y$, $\varphi$, $p_{\rm T}$,  as well as the longitudinal position of the primary vertex (PV) of each event, $z_{\text{vtx}}$. Various selection criteria were applied to minimize residual instrumental effects while optimizing  particle yields. The PV is required to be in the range $|z_{\rm vtx}| \le 6$ cm of the nominal interaction point. 
Tracks are required to have a minimum of 70 reconstructed TPC space points (hits), out of a maximum of 159, and a track fit with $\chi^2$ value per degree of freedom smaller than $2.0$ to ensure good track quality. Contamination of BFs by secondary particles (i.e., weak decays or particles scattered within the detector) is suppressed by requiring  distances of closest approach (DCA) to PV chosen as DCA$_{z} \le 2.0$ cm in the longitudinal direction and DCA$_{xy} \le0.04, 0.04, 2.0$ cm in the transverse plane for $\pi^{\pm}$, ${\rm p/\overline{p}}$, and ${\rm K}^{\pm}$,  respectively. Contamination by $\rm e^+e^-$ pairs from photon conversion is suppressed by removing tracks closer than $1\sigma_{{\rm d}E/{\rm d}x}$ to the TPC Bethe-Bloch median, at a given momentum,  for electrons.

Systematic uncertainties on the amplitudes of $B^{\alpha,\beta}$ and their integrals were calculated as quadratic sums of systematic uncertainties of the correlation function $R_{2}^{\alpha,\beta}$ and the systematic uncertainties on the published single particle densities~\cite{ALICE:2013Pt} used in the computation of the BFs. Uncertainties on $R_{2}^{\alpha,\beta}$   were assessed based on variations of conditions and selection parameters employed in the analysis. A  statistical test~\cite{Barlow:2002yb} was used to identify potential biases introduced by those variations and determine their statistical significance.
Systematic uncertainties, corresponding to a relative deviation  at the maximum of $B^{\alpha,\beta}$  associated with  operation with two solenoidal magnetic field polarities, are smaller than 4\%. Potential biases associated with track selection criteria are up to 3\%, whereas the presence of misidentified and secondary particles contribute up to 4\%, while  kinematic dependencies of the detection efficiency are estimated to be 1\%. Systematic uncertainties on the single particle densities~\cite{ALICE:2013Pt} are species and collision centrality dependent and  typically range from 5 to 10\%.

\begin{figure}[!ht]
 \centering
 \includegraphics[width=0.95\linewidth,trim={1mm 2mm 6mm 5mm},clip]{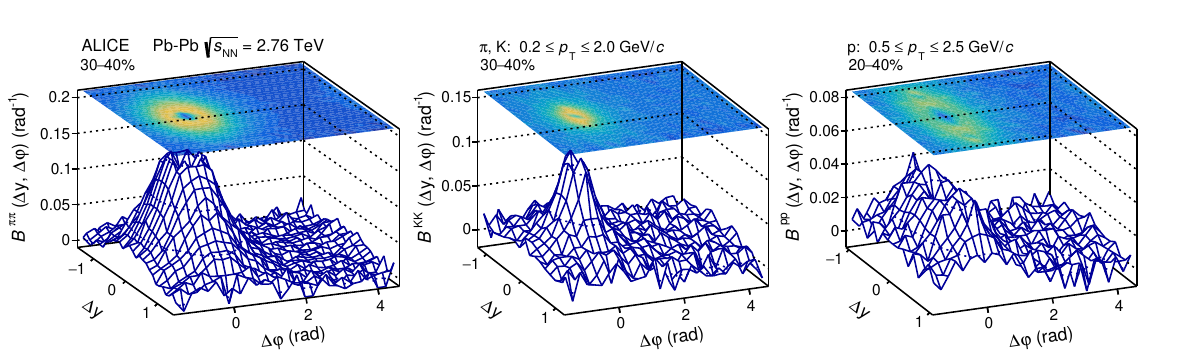}
 \caption{Balance functions $B^{\alpha\beta}(\Delta y,\Delta\varphi)$ of pairs $\alpha\beta = \pi\pi$ (left), ${\rm KK}$ (center), and ${\rm pp}$ (right) measured in semicentral Pb--Pb collisions at $\sqrt{s_{\rm NN}}= 2.76\;\text{TeV}$.}
  \label{fig:2d_BF}
\end{figure}

In order to obtain BF for all nine combinations of $\pi^{\pm}$, ${\rm K}^{\pm}$, and ${\rm p/\overline p}$ species pairs, $R_2^{\alpha\beta}(\Delta y,\Delta\varphi)$ correlators  were first measured, in each centrality class,  for all 36 $\alpha\beta$ permutations of positive and negative $\pi$, ${\rm K}$, and ${\rm p}$. These correlators were then combined according to Eq.~(\ref{eq:BFR2}) and multiplied by the single particle densities $\rho_1^{\beta}$  in the $|y|\le 0.5$ rapidity range~\cite{ALICE:2013Pt}.
Figure~\ref{fig:2d_BF} shows the $B^{\alpha\beta}(\Delta y,\Delta\varphi)$ of $\pi\pi$, ${\rm KK}$, and ${\rm pp}$ pairs in semicentral collisions for illustrative purposes.
The nine measured BFs  exhibit common features, including  prominent near-side peaks centered at $(\Delta y, \Delta \varphi) = (0,0)$ and  relatively flat and featureless away-sides. The flat away-side arises from the fact that positive and negative particles of a given species feature essentially equal azimuthal anisotropy relative to the collision symmetry plane. It is also an indicator of the fast radial flow profile of the  emitting sources, which manifests as strong focusing on the near-side peak~\cite{Voloshin:2006TRE}, although the various species pairs demonstrate different centrality-dependent near-side peak shapes, widths, and magnitudes that indicate that they are subject to different charge balancing pair production and transport mechanisms, as well as final state effects. For instance, $B^{\pi\pi}$ exhibits a deep and narrow dip at $(\Delta y, \Delta \varphi) = (0,0)$, within the near-side correlation peak, resulting in part from the Hanbury Brown--Twiss (HBT) effect, with a depth and width that vary with the source size and thus the centrality~\cite{Jeon:2002BFCF}. $B^{{\rm KK}}$ exhibits much weaker HBT effects, whereas $B^{{\rm pp}}$ also features a narrow dip centered at $(\Delta y, \Delta \varphi) = (0,0)$ within a somewhat elongated near-side peak that may reflect the annihilation of ${\rm p\overline{p}}$ pairs. 
Pairs of protons and antiprotons emitted at small relative $\Delta\eta$ and $\Delta\varphi$ (as well as small relative $p_{\rm T}$) are more likely to interact, and thus annihilate, than pairs produced at large separation, thereby leading to a depletion of pairs near $\Delta y=0$ and $\Delta\varphi=0$.
\begin{figure}[!ht]
 \centering
 \includegraphics[width=0.85\linewidth,trim={0mm 1mm 0mm 0mm},clip]{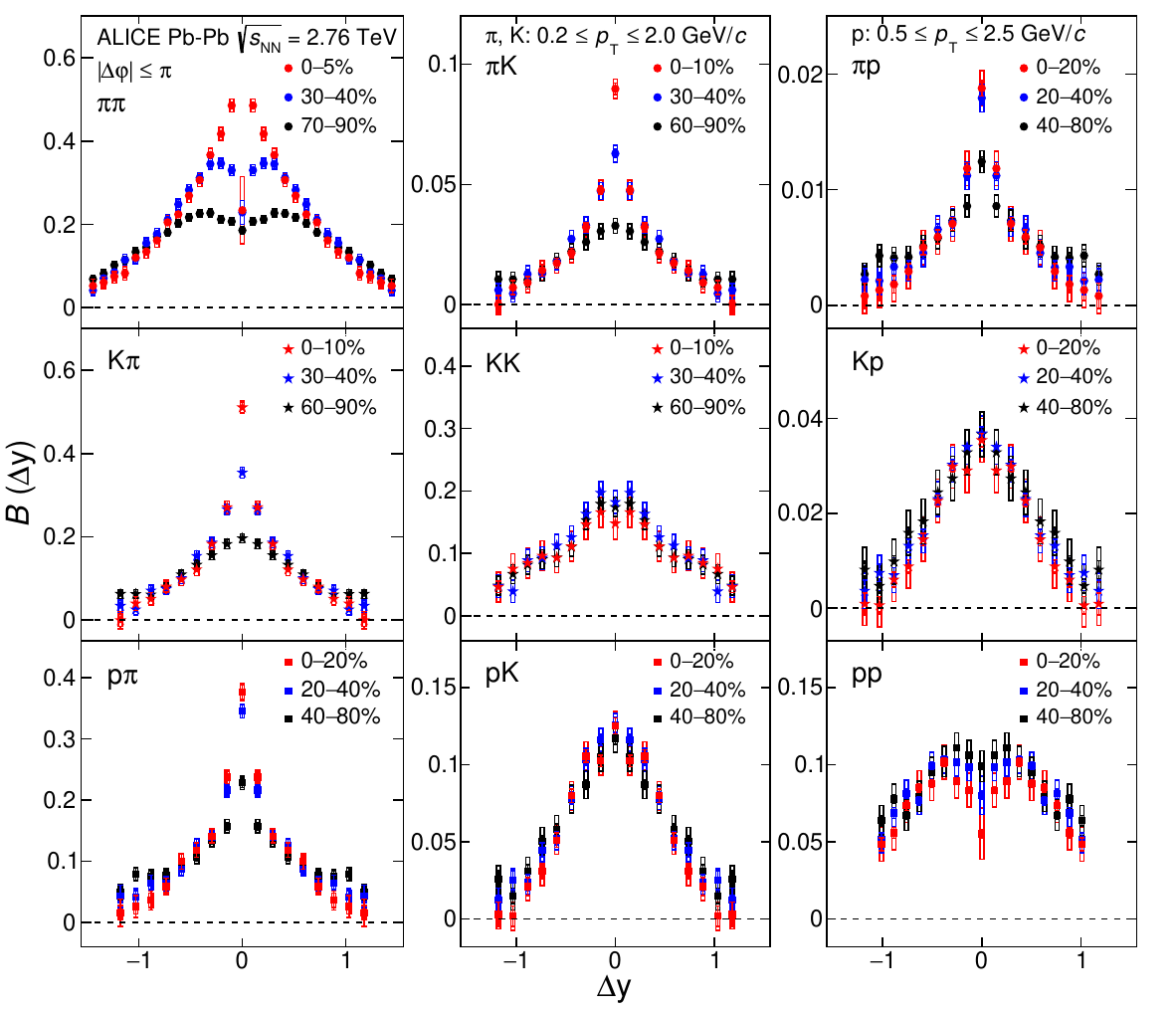}
\caption{Balance function  of species pairs $(\pi,{\rm K},{\rm p})\otimes(\pi,{\rm K},{\rm p})$ projected onto the $\Delta y$  axis for particle pairs within the full range  $|\Delta\varphi|\le\pi$. Vertical bars and open boxes represent statistical and systematic uncertainties, respectively.}
  \label{fig:2}
\end{figure}
\begin{figure}[!ht]
 \centering
 \includegraphics[width=0.85\linewidth,trim={1mm 1mm 1mm 0mm},clip]{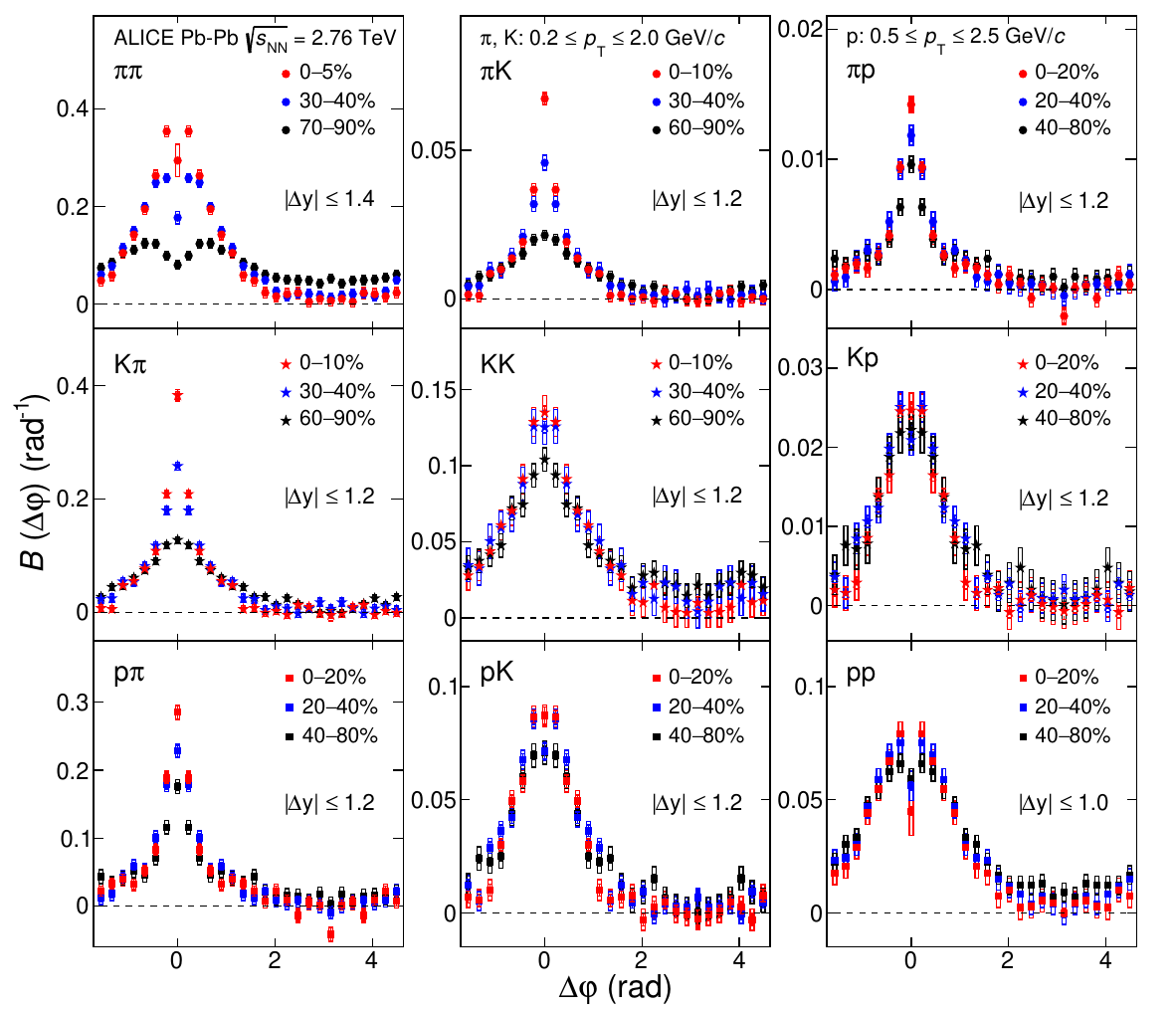}
 \caption{Balance function projections of species pairs $(\pi,{\rm K},{\rm p})\otimes(\pi,{\rm K},{\rm p})$ onto the $\Delta \varphi$  axis for the different particle pairs. Vertical bars and open boxes represent statistical and systematic uncertainties, respectively.}
  \label{fig:3}
\end{figure}

The evolution with collision centrality of $B^{\alpha\beta}$, for all nine combinations $\alpha,\beta=\pi,{\rm K},{\rm p}$, is examined by considering  their projections onto the $\Delta y$ and $\Delta \varphi$ axes in Figs.~\ref{fig:2} and~\ref{fig:3}, respectively. The shape and amplitude of $B^{\pi\pi}$ projections  onto $\Delta y$ exhibit the strongest centrality dependence, whereas those of $B^{\pi {\rm K}}$, $B^{\pi {\rm p}}$, $B^{{\rm K}\pi}$ and $B^{{\rm p}\pi}$ display significantly smaller dependence on centrality. Uncertainties on the rest of the $\Delta y$ projections do not make it possible to claim any centrality dependence albeit some hints are visible in the cases of $B^{{\rm KK}}$ and $B^{{\rm pp}}$. 
\begin{figure}[!ht]
 \centering
 \includegraphics[width=0.95\linewidth]{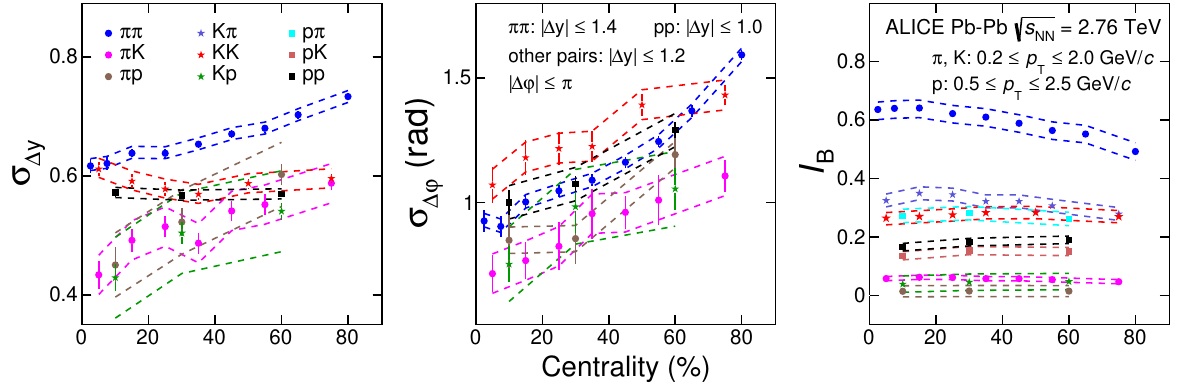}
 \caption{Longitudinal ($\Delta y$) $\sigma$ widths (left), azimuthal ($\Delta\varphi$) $\sigma$ widths (center), and integrals (right) of balance functions of the full species matrix of $\pi^{\pm}$, ${\rm K}^{\pm}$, and ${\rm p/\overline{p}}$ with centrality. For $\Delta y$ and $\Delta\varphi$ widths, ${\rm K}\pi$, ${\rm p}\pi$, and ${\rm pK}$ have the same values with $\pi {\rm K}$, $\pi {\rm p}$, and ${\rm Kp}$, respectively. For the longitudinal widths, the relative azimuthal angle range for all the species pairs is the full azimuth range $|\Delta\varphi|\le\pi$. For the azimuthal widths, the relative rapidity range used for all  species pairs is $|\Delta y|\le1.2$, with the exception of $|\Delta y|\le 1.4$ for $\pi\pi$ and $|\Delta y|\le 1.0$ for $\rm pp$. Vertical bars represent statistical uncertainties while systematic uncertainties are displayed as dash line bands.}
  \label{fig:4}
\end{figure}
The evolution with collision centrality  of the measured BFs is further characterized in terms of their longitudinal and azimuthal standard deviation ($\sigma$) widths, noted $\sigma_{\Delta y}$ and  $\sigma_{\Delta \varphi}$, respectively, as well as their integral, $I_{\rm B}^{\alpha\beta}$, as shown in Fig.~\ref{fig:4}. 
In the longitudinal direction, the widths $\sigma_{\Delta y}$ of all  species pairs, except those of  ${\rm KK}$ and ${\rm pp}$ pairs,  exhibit a significant narrowing from peripheral to central collisions. In contrast, $B^{{\rm KK}}$ is essentially independent in both shape and width  $\sigma_{\Delta y}$ with changing collision centrality, whereas the width $\sigma_{\Delta y}$ of $B^{\rm pp}$ features little centrality dependence even though this balance function exhibits some shape dependence on centrality.

Differences in the evolution of the longitudinal $\sigma$ of pions and kaons BFs were already observed in Au--Au collisions at RHIC~\cite{Aggarwal:2010ya} and were then interpreted as resulting in part from strong radial flow profiles and two-stage emission~\cite{Bass:2000BF,Jeon:2002BFCF}. The independence of the width $\sigma_{\Delta y}$ of the $B^{{\rm KK}}$ relative to the narrowing BFs of all other pairs observed in this work suggests two-stage quark production might also be at play at the TeV collision scale. Indeed, pions might be predominantly formed from the light u, $\rm \bar{u}$, d, and $\rm \bar{d}$ quarks most abundantly produced in the second quark production stage, whereas  kaon production would largely result from   $\rm s\rm \bar{s}$ pairs predominantly created  during the early stages of collisions~\cite{Bass:2000BF,Jeon:2002BFCF}.

Several distinct models have had success in describing the yield of produced hadrons, and more specifically baryons. Such models invoke a range of production mechanisms including parton fragmentation, effective mostly at high-$p_{\rm T}$, as well as  parton coalescence and recombination, playing a predominant role  at low and  intermediate $p_{\rm T}$~\cite{Fries:2003kq,Minissale:2015zwa,Greco:2003xt}. Statistical thermal  models and production models  involving color transparency~\cite{Brodsky:2008qp}  and baryon junctions~\cite{ToporPop:2004lve} have also had a good measure of success. Single particle  spectra of baryons thus do not provide sufficiently discriminating constraints to fully identify baryon production mechanisms.  The added information provided by cross-species BFs shall thus contribute by adding new  constraints for models of particle production and transport. In particular, given that neutrons,  protons,  and their excited states are  composed of light u and d quarks, believed to be  copiously produced in late stage emission (within the context of the two-stage quark production model), it is  conceivable that these baryons are predominantly produced by coalescence (recombination)  of light quarks in the late stage of the collisions. However, baryons (B) and antibaryons ($\overline{\rm B}$) have a relatively large mass and carry a conserved baryonic charge. The question then arises as to whether $\rm B\overline{B}$ correlated  pairs might originate  before the formation of thermalized QGP,  during the early stages of AA  collisions. Late  $\rm B\overline{B}$ production is expected to be characterized by narrow longitudinal BFs while early stage emissions would produce pairs with a much wider $\Delta y$ range~\cite{Bass:2000BF,Jeon:2002BFCF}.
It is clear from Fig.~\ref{fig:2} that $B^{\rm pp}$ must extend  beyond the acceptance of the measurement reported in this paper. This  suggests that pp pairs have rather wide balance functions that might result from early $\rm B\overline B$ pair separation. Detailed models of $\rm B \overline B$ production and transport that account for (strong) decays  from resonant states are required, evidently, to firmly establish this conclusion.

Figure~\ref{fig:4} shows that the $\sigma_{\Delta\varphi}$ widths of the nine BFs exhibit narrowing trends from peripheral to central collisions. The widths $\sigma_{\Delta\varphi}$ feature a wide spread of values at a given collision centrality, with those of ${\rm KK}$ pairs being the largest and those of ${\rm \pi K}$ the smallest. The widths also exhibit similar reductions with increasing collision centrality.
These observations  are in agreement  with azimuthal BFs already reported from observations at RHIC for unidentified charged particle and identified $\pi\pi$, $\rm KK$ pairs~\cite{Adams:2003kg,Aggarwal:2010ya}, as well as unidentified charged particle BFs in collisions at the LHC~\cite{Abelev:2013csa,Adam:2015gda}.
This narrowing can be qualitatively understood as resulting from the larger estimated transverse expansion velocity present in more central AA  collisions~\cite{ALICE:2013mez}.
It competes with an 
opposing trend associated with light quark diffusivity, expected to broaden and  smear out the long range  tails of the $\Delta\varphi$ BFs for systems featuring increasingly large lifespans~\cite{Pratt:2019pnd}. Given the radial boost profile and contributions from resonance decays can be largely calibrated based on the shape of single particle $p_{\rm T}$ spectra, the BF projections presented in Fig.~\ref{fig:2}, ~\ref{fig:3} and the evolution of their widths $\sigma_{\Delta y}$ and $\sigma_{\Delta\varphi}$, shown in Fig.~\ref{fig:4}, then provide the first comprehensive set of azimuthal BFs to estimate the diffusivity of light quarks at the LHC~\cite{Pratt:2019pnd, Pratt:2018ebf}.
The above discussion  neglects possible contributions from the fragmentation of jets but these are anticipated to be small in the $p_{\rm T}$ range of this measurement. 
Quantitative estimates of such contributions would  need to be accounted for in theoretical modeling of balance functions reported in this work for the purpose of determining the diffusivity of light quarks. 

Contributions of $\phi \rightarrow {\rm K}^+ +{ \rm K}^-$ decays to $B^{\rm KK}$ were studied using simulated events from the HIJING generator~\cite{Wang:1991hta}. The amplitude of the near-side peak of  $B^{\rm KK}$ is reduced by about 30\% when contributions from $\phi$-meson decays are explicitly excluded, while the correlator $\Delta y$ and $\Delta\varphi$ widths increase by about 7--8\%. Effects associated with radial flow, not present in HIJING, could reduce this broadening effect and possibly induce a narrowing of the $\Delta y$ width of $B^{\rm KK}$ in more central collisions. However, no such narrowing is observed thereby signaling a more intricate production and transport evolution with competing contributions from $\phi$ produced at hadronization of the QGP and by coalescence of kaons within a hadron phase.

The evolution  with the collision centrality of the integrals $I_{B}^{\alpha\beta}$ of the  nine species-pairs $B^{\alpha\beta}(\Delta y,\Delta\varphi)$ shown  in the right panel of Fig.~\ref{fig:4}  
is also of considerable interest.
By definition, a balance function $B^{\alpha\beta}(\Delta y,\Delta\varphi, \Delta p_{\rm T})$ measures the ``likelihood" of finding a charge balancing  particle of a type $\beta$, e.g., $\pi^+$, with a pair separation $\Delta y$, $\Delta\varphi$, $\Delta p_{\rm T}$ away from a reference particle of type $\alpha$, e.g., $\pi^-$.
But charge balancing can be accomplished, on average, by distinct species, e.g., $\rm p$, $\rm K^+$, and more rarely produced heavier particles, in additions to $\pi^+$.
The integral, $I^{\alpha\beta}_{B}(4\pi)$, of $B^{\alpha\beta}(\Delta y,\Delta\varphi, \Delta p_{\rm T})$  over the full phase space   is thus proportional to the average fraction (and probability in the full phase space limit) of balancing partners of species $\beta$.
Indeed, neglecting contributions from species other than pions, kaons, and protons, one expects the sum,  $I^{\alpha}_{B}(4\pi) \equiv I^{\alpha\pi}_{B}(4\pi)+I^{\alpha\rm K}_{B}(4\pi)+I^{\alpha\rm p}_{B}(4\pi)$ to converge to unity, $I^{\alpha}_{B}(4\pi)\approx 1$, in the full acceptance limit~\cite{Pruneau:2019BNC}.
Integrals $I^{\alpha\beta}_{B}(4\pi)$ thus amount to probabilities $I^{\alpha\beta}_{B}(4\pi)/I^{\alpha}_{B}(4\pi)$ of having charge balancing of a species $\alpha$ by a species of type $\beta$ and are  indicators of the hadronization chemistry of the QGP, that is, what fraction of species $\alpha$  are accompanied (balanced), on average, by a species $\beta$~\cite{Pruneau:2019BNC}.
However, when measured in a limited acceptance, integrals $I^{\alpha\beta}_{B}(\Omega<4\pi)$ cannot, strictly speaking,   be considered charge balancing probabilities. They nonetheless provide useful indicators of the hadrochemistry as well as the  flavor and baryon number transport in AA  collisions. As such, integrals $I^{\alpha\beta}_{B}$ shown in Fig.~\ref{fig:4} as a function of collision centrality are surprising on two accounts. First, they show that the  balance fractions are all, but one, approximately independent of collision centrality. The notable exception is the $\pi\pi$ integral which increases by about 20\% from peripheral to central collisions.
Second, close examination of these pairing fractions shows they are rather different than inclusive probabilities of observing $\pi$, ${\rm K}$, and ${\rm p/\overline p}$ in Pb--Pb collisions. For instance, $I_{\rm B}^{{\rm K}\pi}$ is not larger than $I_{\rm B}^{\rm KK}$ by the $\pi/{\rm K}\sim 6.7$ ratio of inclusive single particle yields and $I_{\rm B}^{\rm pp}$ is larger than $I_{\rm B}^{\rm pK}$ also in contrast to observed  ${\rm K/p}\sim 3$ yield ratios~\cite{ALICE:2013Pt}. Hadron species charge balancing pairing fractions are thus indeed very different than the relative probabilities of single hadrons, and as such, provide new and useful information to further probe the hadronization of the QGP.
This difference arises because the set of processes $\mathcal{P}_2$ that lead to a specific balancing pair $\alpha\beta$ (e.g., $\mathcal{P}_2: \rightarrow \alpha^{\pm} + \beta^{\mp} + X$) is, by construction, far smaller than the set of processes $\mathcal{P}_1$ leading to a given particle species $\alpha$ or $\beta$ (e.g., $\mathcal{P}_1: \rightarrow \alpha^{\pm} + X$ or $\mathcal{P}_1: \rightarrow \beta^{\mp} + Y$). It is remarkable, nonetheless, that the pairing fractions $I_{\rm B}^{\alpha\beta}$ exhibit essentially no collision centrality dependence while single particle yield ratios are known to exhibit a weak dependence on collision centrality~\cite{STAR:2007zea,ALICE:2019hno}. Note that the observed rise of $I_{\rm B}^{\pi\pi}$ in more central collisions may artificially result from increased kinematic focusing of pions with centrality in the  $p_{\rm T}$ and $\Delta y$ acceptance of this measurement. The higher velocity flow fields encountered in more central Pb--Pb collisions could indeed shift and focus the  yield of associated pions. Why such a shift is not as important for other charge balancing pairs remains to be elucidated with a comprehensive model accounting for the  flow velocity profile and appropriate sets of charge conserving processes yielding balancing charges in the final state of collisions.  Recent deployments of hydrodynamic models feature the former but lack the latter~\cite{Oliinychenko:2020cmr,Vovchenko:2020kwg,Oliinychenko:2019zfk}. Further theoretical work is thus required to interpret the observed collision centrality dependence of the pairing probabilities displayed in Fig.~\ref{fig:4}. As such calculations become available, the data reported in this work, and specifically the integral $I^{\alpha\beta}_B$ shown in Fig.~\ref{fig:4}, shall provide increased sensitivity to the hadrochemistry of the QGP and its susceptibilities. 

In summary, this paper presents the first measurements of the collision centrality evolution of same and cross-species balance functions of identified $\pi^{\pm}$, ${\rm K}^{\pm}$ and $\rm p/\overline p$ at the LHC. Measured as functions of particle pair separation in  rapidity ($\Delta y$) and azimuth ($\Delta\varphi$), the BFs exhibit  prominent near-side peaks centered at $(\Delta y,\Delta\varphi)=(0,0)$ which feature different shapes, amplitudes, and widths, and varied dependencies on collision centrality. The BFs of species-pairs measured in this work feature narrowing $\Delta\varphi$ widths in more central collisions, owing to the strong radial flow field present in central Pb--Pb collisions. Theoretical studies beyond the scope of this work shall use this data to put upper limits  on the diffusivity coefficients of light quarks. In the longitudinal direction, the $\sigma$ widths of BFs of all species pairs decrease with centrality except for those of ${\rm KK}$ and ${\rm pp}$ pairs. The shape and width of ${\rm KK}$ BFs are independent of collision centrality, while the ${\rm pp}$ BFs peak shapes depend only minimally on centrality. The observed centrality independence of the ${\rm KK}$ and narrowing $\sigma$ of other species in the longitudinal direction are  qualitatively consistent with effects associated with radial flow and the two-stage quark production scenario, which posits that quark production occurs predominantly in early and late stages separated by a period of isentropic expansion.
Integrals $I^{\alpha\beta}_B$ 
constitute an important finding of this study as they indicate that pairing fractions $I^{\alpha\beta}_B$ are nearly independent of  collision centrality, and provide a valuable quantitative characterization  of the hadronization of the QGP.


\newenvironment{acknowledgement}{\relax}{\relax}
\begin{acknowledgement}
\section*{Acknowledgements}

The ALICE Collaboration would like to thank all its engineers and technicians for their invaluable contributions to the construction of the experiment and the CERN accelerator teams for the outstanding performance of the LHC complex.
The ALICE Collaboration gratefully acknowledges the resources and support provided by all Grid centres and the Worldwide LHC Computing Grid (WLCG) collaboration.
The ALICE Collaboration acknowledges the following funding agencies for their support in building and running the ALICE detector:
A. I. Alikhanyan National Science Laboratory (Yerevan Physics Institute) Foundation (ANSL), State Committee of Science and World Federation of Scientists (WFS), Armenia;
Austrian Academy of Sciences, Austrian Science Fund (FWF): [M 2467-N36] and Nationalstiftung f\"{u}r Forschung, Technologie und Entwicklung, Austria;
Ministry of Communications and High Technologies, National Nuclear Research Center, Azerbaijan;
Conselho Nacional de Desenvolvimento Cient\'{\i}fico e Tecnol\'{o}gico (CNPq), Financiadora de Estudos e Projetos (Finep), Funda\c{c}\~{a}o de Amparo \`{a} Pesquisa do Estado de S\~{a}o Paulo (FAPESP) and Universidade Federal do Rio Grande do Sul (UFRGS), Brazil;
Ministry of Education of China (MOEC) , Ministry of Science \& Technology of China (MSTC) and National Natural Science Foundation of China (NSFC), China;
Ministry of Science and Education and Croatian Science Foundation, Croatia;
Centro de Aplicaciones Tecnol\'{o}gicas y Desarrollo Nuclear (CEADEN), Cubaenerg\'{\i}a, Cuba;
Ministry of Education, Youth and Sports of the Czech Republic, Czech Republic;
The Danish Council for Independent Research | Natural Sciences, the VILLUM FONDEN and Danish National Research Foundation (DNRF), Denmark;
Helsinki Institute of Physics (HIP), Finland;
Commissariat \`{a} l'Energie Atomique (CEA) and Institut National de Physique Nucl\'{e}aire et de Physique des Particules (IN2P3) and Centre National de la Recherche Scientifique (CNRS), France;
Bundesministerium f\"{u}r Bildung und Forschung (BMBF) and GSI Helmholtzzentrum f\"{u}r Schwerionenforschung GmbH, Germany;
General Secretariat for Research and Technology, Ministry of Education, Research and Religions, Greece;
National Research, Development and Innovation Office, Hungary;
Department of Atomic Energy Government of India (DAE), Department of Science and Technology, Government of India (DST), University Grants Commission, Government of India (UGC) and Council of Scientific and Industrial Research (CSIR), India;
Indonesian Institute of Science, Indonesia;
Istituto Nazionale di Fisica Nucleare (INFN), Italy;
Japanese Ministry of Education, Culture, Sports, Science and Technology (MEXT), Japan Society for the Promotion of Science (JSPS) KAKENHI and Japanese Ministry of Education, Culture, Sports, Science and Technology (MEXT)of Applied Science (IIST), Japan;
Consejo Nacional de Ciencia (CONACYT) y Tecnolog\'{i}a, through Fondo de Cooperaci\'{o}n Internacional en Ciencia y Tecnolog\'{i}a (FONCICYT) and Direcci\'{o}n General de Asuntos del Personal Academico (DGAPA), Mexico;
Nederlandse Organisatie voor Wetenschappelijk Onderzoek (NWO), Netherlands;
The Research Council of Norway, Norway;
Commission on Science and Technology for Sustainable Development in the South (COMSATS), Pakistan;
Pontificia Universidad Cat\'{o}lica del Per\'{u}, Peru;
Ministry of Education and Science, National Science Centre and WUT ID-UB, Poland;
Korea Institute of Science and Technology Information and National Research Foundation of Korea (NRF), Republic of Korea;
Ministry of Education and Scientific Research, Institute of Atomic Physics and Ministry of Research and Innovation and Institute of Atomic Physics, Romania;
Joint Institute for Nuclear Research (JINR), Ministry of Education and Science of the Russian Federation, National Research Centre Kurchatov Institute, Russian Science Foundation and Russian Foundation for Basic Research, Russia;
Ministry of Education, Science, Research and Sport of the Slovak Republic, Slovakia;
National Research Foundation of South Africa, South Africa;
Swedish Research Council (VR) and Knut \& Alice Wallenberg Foundation (KAW), Sweden;
European Organization for Nuclear Research, Switzerland;
Suranaree University of Technology (SUT), National Science and Technology Development Agency (NSDTA) and Office of the Higher Education Commission under NRU project of Thailand, Thailand;
Turkish Energy, Nuclear and Mineral Research Agency (TENMAK), Turkey;
National Academy of  Sciences of Ukraine, Ukraine;
Science and Technology Facilities Council (STFC), United Kingdom;
National Science Foundation of the United States of America (NSF) and United States Department of Energy, Office of Nuclear Physics (DOE NP), United States of America.    
\end{acknowledgement}

\bibliographystyle{utphys}   
\bibliography{main}

\newpage
\appendix

%
%

\section{The ALICE Collaboration}
\label{app:collab}
\small
\begin{flushleft} 

S.~Acharya$^{\rm 142}$, 
D.~Adamov\'{a}$^{\rm 97}$, 
A.~Adler$^{\rm 75}$, 
J.~Adolfsson$^{\rm 82}$, 
G.~Aglieri Rinella$^{\rm 34}$, 
M.~Agnello$^{\rm 30}$, 
N.~Agrawal$^{\rm 54}$, 
Z.~Ahammed$^{\rm 142}$, 
S.~Ahmad$^{\rm 16}$, 
S.U.~Ahn$^{\rm 77}$
% ...

\section*{Affiliation notes}

$^{\rm I}$ Deceased\\
$^{\rm II}$ Also at: Italian National Agency for New Technologies, Energy and Sustainable Economic Development (ENEA), Bologna, Italy\\
$^{\rm III}$ Also at: Dipartimento DET del Politecnico di Torino, Turin, Italy\\
$^{\rm IV}$ Also at: M.V. Lomonosov Moscow State University, D.V. Skobeltsyn Institute of Nuclear, Physics, Moscow, Russia\\
$^{\rm V}$ Also at: Department of Applied Physics, Aligarh Muslim University, Aligarh, India
\\
$^{\rm VI}$ Also at: Institute of Theoretical Physics, University of Wroclaw, Poland\\
$^{\rm VII}$ Also at: University of Kansas, Lawrence, Kansas, United States\\

\section*{Collaboration Institutes}

$^{1}$ A.I. Alikhanyan National Science Laboratory (Yerevan Physics Institute) Foundation, Yerevan, Armenia\\
$^{2}$ AGH University of Science and Technology, Cracow, Poland\\
$^{3}$ Bogolyubov Institute for Theoretical Physics, National Academy of Sciences of Ukraine, Kiev, Ukraine\\
$^{4}$ Bose Institute, Department of Physics  and Centre for Astroparticle Physics and Space Science (CAPSS), Kolkata, India\\
$^{5}$ Budker Institute for Nuclear Physics, Novosibirsk, Russia\\
$^{6}$ California Polytechnic State University, San Luis Obispo, California, United States\\
$^{7}$ Central China Normal University, Wuhan, China\\
$^{8}$ Centro de Aplicaciones Tecnol\'{o}gicas y Desarrollo Nuclear (CEADEN), Havana, Cuba\\
$^{9}$ Centro de Investigaci\'{o}n y de Estudios Avanzados (CINVESTAV), Mexico City and M\'{e}rida, Mexico\\
$^{10}$ Chicago State University, Chicago, Illinois, United States\\
$^{11}$ China Institute of Atomic Energy, Beijing, China\\
$^{12}$ Chungbuk National University, Cheongju, Republic of Korea\\
$^{13}$ Comenius University Bratislava, Faculty of Mathematics, Physics and Informatics, Bratislava, Slovakia\\
$^{14}$ COMSATS University Islamabad, Islamabad, Pakistan\\
$^{15}$ Creighton University, Omaha, Nebraska, United States\\
$^{16}$ Department of Physics, Aligarh Muslim University, Aligarh, India\\
$^{17}$ Department of Physics, Pusan National University, Pusan, Republic of Korea\\
$^{18}$ Department of Physics, Sejong University, Seoul, Republic of Korea\\
$^{19}$ Department of Physics, University of California, Berkeley, California, United States\\
$^{20}$ Department of Physics, University of Oslo, Oslo, Norway\\
$^{21}$ Department of Physics and Technology, University of Bergen, Bergen, Norway\\
$^{22}$ Dipartimento di Fisica dell'Universit\`{a} and Sezione INFN, Cagliari, Italy\\
$^{23}$ Dipartimento di Fisica dell'Universit\`{a} and Sezione INFN, Trieste, Italy\\
$^{24}$ Dipartimento di Fisica dell'Universit\`{a} and Sezione INFN, Turin, Italy\\
$^{25}$ Dipartimento di Fisica e Astronomia dell'Universit\`{a} and Sezione INFN, Bologna, Italy\\
$^{26}$ Dipartimento di Fisica e Astronomia dell'Universit\`{a} and Sezione INFN, Catania, Italy\\
$^{27}$ Dipartimento di Fisica e Astronomia dell'Universit\`{a} and Sezione INFN, Padova, Italy\\
$^{28}$ Dipartimento di Fisica e Nucleare e Teorica, Universit\`{a} di Pavia, Pavia, Italy\\
$^{29}$ Dipartimento di Fisica `E.R.~Caianiello' dell'Universit\`{a} and Gruppo Collegato INFN, Salerno, Italy\\
$^{30}$ Dipartimento DISAT del Politecnico and Sezione INFN, Turin, Italy\\
$^{31}$ Dipartimento di Scienze e Innovazione Tecnologica dell'Universit\`{a} del Piemonte Orientale and INFN Sezione di Torino, Alessandria, Italy\\
$^{32}$ Dipartimento di Scienze MIFT, Universit\`{a} di Messina, Messina, Italy\\
$^{33}$ Dipartimento Interateneo di Fisica `M.~Merlin' and Sezione INFN, Bari, Italy\\
$^{34}$ European Organization for Nuclear Research (CERN), Geneva, Switzerland\\
$^{35}$ Faculty of Electrical Engineering, Mechanical Engineering and Naval Architecture, University of Split, Split, Croatia\\
$^{36}$ Faculty of Engineering and Science, Western Norway University of Applied Sciences, Bergen, Norway\\
$^{37}$ Faculty of Nuclear Sciences and Physical Engineering, Czech Technical University in Prague, Prague, Czech Republic\\
$^{38}$ Faculty of Science, P.J.~\v{S}af\'{a}rik University, Ko\v{s}ice, Slovakia\\
$^{39}$ Frankfurt Institute for Advanced Studies, Johann Wolfgang Goethe-Universit\"{a}t Frankfurt, Frankfurt, Germany\\
$^{40}$ Fudan University, Shanghai, China\\
$^{41}$ Gangneung-Wonju National University, Gangneung, Republic of Korea\\
$^{42}$ Gauhati University, Department of Physics, Guwahati, India\\
$^{43}$ Helmholtz-Institut f\"{u}r Strahlen- und Kernphysik, Rheinische Friedrich-Wilhelms-Universit\"{a}t Bonn, Bonn, Germany\\
$^{44}$ Helsinki Institute of Physics (HIP), Helsinki, Finland\\
$^{45}$ High Energy Physics Group,  Universidad Aut\'{o}noma de Puebla, Puebla, Mexico\\
$^{46}$ Hiroshima University, Hiroshima, Japan\\
$^{47}$ Hochschule Worms, Zentrum  f\"{u}r Technologietransfer und Telekommunikation (ZTT), Worms, Germany\\
$^{48}$ Horia Hulubei National Institute of Physics and Nuclear Engineering, Bucharest, Romania\\
$^{49}$ Indian Institute of Technology Bombay (IIT), Mumbai, India\\
$^{50}$ Indian Institute of Technology Indore, Indore, India\\
$^{51}$ Indonesian Institute of Sciences, Jakarta, Indonesia\\
$^{52}$ INFN, Laboratori Nazionali di Frascati, Frascati, Italy\\
$^{53}$ INFN, Sezione di Bari, Bari, Italy\\
$^{54}$ INFN, Sezione di Bologna, Bologna, Italy\\
$^{55}$ INFN, Sezione di Cagliari, Cagliari, Italy\\
$^{56}$ INFN, Sezione di Catania, Catania, Italy\\
$^{57}$ INFN, Sezione di Padova, Padova, Italy\\
$^{58}$ INFN, Sezione di Pavia, Pavia, Italy\\
$^{59}$ INFN, Sezione di Roma, Rome, Italy\\
$^{60}$ INFN, Sezione di Torino, Turin, Italy\\
$^{61}$ INFN, Sezione di Trieste, Trieste, Italy\\
$^{62}$ Inha University, Incheon, Republic of Korea\\
$^{63}$ Institute for Gravitational and Subatomic Physics (GRASP), Utrecht University/Nikhef, Utrecht, Netherlands\\
$^{64}$ Institute for Nuclear Research, Academy of Sciences, Moscow, Russia\\
$^{65}$ Institute of Experimental Physics, Slovak Academy of Sciences, Ko\v{s}ice, Slovakia\\
$^{66}$ Institute of Physics, Homi Bhabha National Institute, Bhubaneswar, India\\
$^{67}$ Institute of Physics of the Czech Academy of Sciences, Prague, Czech Republic\\
$^{68}$ Institute of Space Science (ISS), Bucharest, Romania\\
$^{69}$ Institut f\"{u}r Kernphysik, Johann Wolfgang Goethe-Universit\"{a}t Frankfurt, Frankfurt, Germany\\
$^{70}$ Instituto de Ciencias Nucleares, Universidad Nacional Aut\'{o}noma de M\'{e}xico, Mexico City, Mexico\\
$^{71}$ Instituto de F\'{i}sica, Universidade Federal do Rio Grande do Sul (UFRGS), Porto Alegre, Brazil\\
$^{72}$ Instituto de F\'{\i}sica, Universidad Nacional Aut\'{o}noma de M\'{e}xico, Mexico City, Mexico\\
$^{73}$ iThemba LABS, National Research Foundation, Somerset West, South Africa\\
$^{74}$ Jeonbuk National University, Jeonju, Republic of Korea\\
$^{75}$ Johann-Wolfgang-Goethe Universit\"{a}t Frankfurt Institut f\"{u}r Informatik, Fachbereich Informatik und Mathematik, Frankfurt, Germany\\
$^{76}$ Joint Institute for Nuclear Research (JINR), Dubna, Russia\\
$^{77}$ Korea Institute of Science and Technology Information, Daejeon, Republic of Korea\\
$^{78}$ KTO Karatay University, Konya, Turkey\\
$^{79}$ Laboratoire de Physique des 2 Infinis, Ir\`{e}ne Joliot-Curie, Orsay, France\\
$^{80}$ Laboratoire de Physique Subatomique et de Cosmologie, Universit\'{e} Grenoble-Alpes, CNRS-IN2P3, Grenoble, France\\
$^{81}$ Lawrence Berkeley National Laboratory, Berkeley, California, United States\\
$^{82}$ Lund University Department of Physics, Division of Particle Physics, Lund, Sweden\\
$^{83}$ Moscow Institute for Physics and Technology, Moscow, Russia\\
$^{84}$ Nagasaki Institute of Applied Science, Nagasaki, Japan\\
$^{85}$ Nara Women{'}s University (NWU), Nara, Japan\\
$^{86}$ National and Kapodistrian University of Athens, School of Science, Department of Physics , Athens, Greece\\
$^{87}$ National Centre for Nuclear Research, Warsaw, Poland\\
$^{88}$ National Institute of Science Education and Research, Homi Bhabha National Institute, Jatni, India\\
$^{89}$ National Nuclear Research Center, Baku, Azerbaijan\\
$^{90}$ National Research Centre Kurchatov Institute, Moscow, Russia\\
$^{91}$ Niels Bohr Institute, University of Copenhagen, Copenhagen, Denmark\\
$^{92}$ Nikhef, National institute for subatomic physics, Amsterdam, Netherlands\\
$^{93}$ NRC Kurchatov Institute IHEP, Protvino, Russia\\
$^{94}$ NRC \guillemotleft Kurchatov\guillemotright  Institute - ITEP, Moscow, Russia\\
$^{95}$ NRNU Moscow Engineering Physics Institute, Moscow, Russia\\
$^{96}$ Nuclear Physics Group, STFC Daresbury Laboratory, Daresbury, United Kingdom\\
$^{97}$ Nuclear Physics Institute of the Czech Academy of Sciences, \v{R}e\v{z} u Prahy, Czech Republic\\
$^{98}$ Oak Ridge National Laboratory, Oak Ridge, Tennessee, United States\\
$^{99}$ Ohio State University, Columbus, Ohio, United States\\
$^{100}$ Petersburg Nuclear Physics Institute, Gatchina, Russia\\
$^{101}$ Physics department, Faculty of science, University of Zagreb, Zagreb, Croatia\\
$^{102}$ Physics Department, Panjab University, Chandigarh, India\\
$^{103}$ Physics Department, University of Jammu, Jammu, India\\
$^{104}$ Physics Department, University of Rajasthan, Jaipur, India\\
$^{105}$ Physikalisches Institut, Eberhard-Karls-Universit\"{a}t T\"{u}bingen, T\"{u}bingen, Germany\\
$^{106}$ Physikalisches Institut, Ruprecht-Karls-Universit\"{a}t Heidelberg, Heidelberg, Germany\\
$^{107}$ Physik Department, Technische Universit\"{a}t M\"{u}nchen, Munich, Germany\\
$^{108}$ Politecnico di Bari and Sezione INFN, Bari, Italy\\
$^{109}$ Research Division and ExtreMe Matter Institute EMMI, GSI Helmholtzzentrum f\"ur Schwerionenforschung GmbH, Darmstadt, Germany\\
$^{110}$ Russian Federal Nuclear Center (VNIIEF), Sarov, Russia\\
$^{111}$ Saha Institute of Nuclear Physics, Homi Bhabha National Institute, Kolkata, India\\
$^{112}$ School of Physics and Astronomy, University of Birmingham, Birmingham, United Kingdom\\
$^{113}$ Secci\'{o}n F\'{\i}sica, Departamento de Ciencias, Pontificia Universidad Cat\'{o}lica del Per\'{u}, Lima, Peru\\
$^{114}$ St. Petersburg State University, St. Petersburg, Russia\\
$^{115}$ Stefan Meyer Institut f\"{u}r Subatomare Physik (SMI), Vienna, Austria\\
$^{116}$ SUBATECH, IMT Atlantique, Universit\'{e} de Nantes, CNRS-IN2P3, Nantes, France\\
$^{117}$ Suranaree University of Technology, Nakhon Ratchasima, Thailand\\
$^{118}$ Technical University of Ko\v{s}ice, Ko\v{s}ice, Slovakia\\
$^{119}$ The Henryk Niewodniczanski Institute of Nuclear Physics, Polish Academy of Sciences, Cracow, Poland\\
$^{120}$ The University of Texas at Austin, Austin, Texas, United States\\
$^{121}$ Universidad Aut\'{o}noma de Sinaloa, Culiac\'{a}n, Mexico\\
$^{122}$ Universidade de S\~{a}o Paulo (USP), S\~{a}o Paulo, Brazil\\
$^{123}$ Universidade Estadual de Campinas (UNICAMP), Campinas, Brazil\\
$^{124}$ Universidade Federal do ABC, Santo Andre, Brazil\\
$^{125}$ University of Cape Town, Cape Town, South Africa\\
$^{126}$ University of Houston, Houston, Texas, United States\\
$^{127}$ University of Jyv\"{a}skyl\"{a}, Jyv\"{a}skyl\"{a}, Finland\\
$^{128}$ University of Kansas, Lawrence, Kansas, United States\\
$^{129}$ University of Liverpool, Liverpool, United Kingdom\\
$^{130}$ University of Science and Technology of China, Hefei, China\\
$^{131}$ University of South-Eastern Norway, Tonsberg, Norway\\
$^{132}$ University of Tennessee, Knoxville, Tennessee, United States\\
$^{133}$ University of the Witwatersrand, Johannesburg, South Africa\\
$^{134}$ University of Tokyo, Tokyo, Japan\\
$^{135}$ University of Tsukuba, Tsukuba, Japan\\
$^{136}$ Universit\'{e} Clermont Auvergne, CNRS/IN2P3, LPC, Clermont-Ferrand, France\\
$^{137}$ Universit\'{e} de Lyon, CNRS/IN2P3, Institut de Physique des 2 Infinis de Lyon, Lyon, France\\
$^{138}$ Universit\'{e} de Strasbourg, CNRS, IPHC UMR 7178, F-67000 Strasbourg, France, Strasbourg, France\\
$^{139}$ Universit\'{e} Paris-Saclay Centre d'Etudes de Saclay (CEA), IRFU, D\'{e}partment de Physique Nucl\'{e}aire (DPhN), Saclay, France\\
$^{140}$ Universit\`{a} degli Studi di Foggia, Foggia, Italy\\
$^{141}$ Universit\`{a} di Brescia, Brescia, Italy\\
$^{142}$ Variable Energy Cyclotron Centre, Homi Bhabha National Institute, Kolkata, India\\
$^{143}$ Warsaw University of Technology, Warsaw, Poland\\
$^{144}$ Wayne State University, Detroit, Michigan, United States\\
$^{145}$ Westf\"{a}lische Wilhelms-Universit\"{a}t M\"{u}nster, Institut f\"{u}r Kernphysik, M\"{u}nster, Germany\\
$^{146}$ Wigner Research Centre for Physics, Budapest, Hungary\\
$^{147}$ Yale University, New Haven, Connecticut, United States\\
$^{148}$ Yonsei University, Seoul, Republic of Korea\\

\end{flushleft} 
\end{document}